\documentclass[modern]{aastex631}
\usepackage{xspace}

\newcommand\msun{\ensuremath{M_\odot}\xspace}

\newcommand\avgamp{\ensuremath{\left < \, \mathrm{A} \, \right >}\xspace}

\shorttitle{Rapid Rotation of EGGR 156}
\shortauthors{Williams et al.}
\accepted{27 July 2022}
\submitjournal{The Astronomical Journal}

\graphicspath{{./}{figures/}}

\begin{document}

\title{The Rapid Rotation of the Strongly Magnetic Ultramassive White Dwarf EGGR 156}

\author[0000-0002-1413-7679]{Kurtis A.\ Williams}
\affiliation{Department of Physics \& Astronomy \\
Texas A\&M University-Commerce  \\
P.O. Box 3011 \\
Commerce, TX 75429-3011, USA}

\author[0000-0001-5941-2286]{J.\ J.\ Hermes}
\affiliation{Department of Astronomy \& Institute for Astrophysical Research\\ 
Boston University\\ 
Boston, MA 02215, USA}

\author[0000-0002-0853-3464]{Zachary P.\ Vanderbosch}
\affiliation{Division of Physics, Mathematics, and Astronomy\\ 
California Institute of Technology\\ 
Pasadena, CA 91125, USA}


\begin{abstract}
The distribution of white dwarf rotation periods provides a means for constraining angular momentum evolution during the late stages of stellar evolution, as well as insight into the physics and remnants of double degenerate mergers.  Although the rotational distribution of low mass white dwarfs is relatively well constrained via asteroseismology, that of high mass white dwarfs, which can arise from either intermediate mass stellar evolution or white dwarf mergers, is not.  Photometric variability in white dwarfs due to rotation of a spotted star is rapidly increasing the sample size of high mass white dwarfs with measured rotation periods.  We present the discovery of 22.4 minute photometric variability in the lightcurve of EGGR 156, a strongly magnetic, ultramassive white dwarf.  We interpret this variability as rapid rotation, and our data suggest that EGGR 156 is the remnant of a double degenerate merger. Finally, we calculate the rate of period change in rapidly rotating, massive, magnetic WDs due to magnetic dipole radiation.  In many cases, including EGGR 156, the period change is not currently detectable over reasonable timescales, indicating that these WDs could be very precise clocks.  For the most highly magnetic, rapidly rotating massive WDs, such as ZTF J1901+1450 and RE J0317$-$853, the period change should be detectable and may help constrain the structure and evolution of these exotic white dwarfs.  
\end{abstract}

\section{Introduction} \label{sec.intro}

The distribution of rotational speeds of white dwarfs (WDs) can be used to constrain the core-envelope coupling in asymptotic giant branch (AGB) stars, assuming single star evolution \citep[e.g.,][]{2004IAUS..215..561K,2015ASPC..493...65K,2019ARA&A..57...35A}.  This coupling is extraordinarily difficult to predict from first principles, though informed arguments suggest that low-mass stars should produce WDs with rotation periods $\gtrsim 5$ hr, while those of intermediate masses might rotate with periods as short as $\approx 400$ s \citep[e.g.,][]{2013ApJ...775L...1T,2015ASPC..493...65K}. 

Observational constraints on core rotation rates in post-main sequence stages of evolution are available via asteroseismic observations of evolved stars such as red giants and core He-burning giant branch stars \citep[e.g.,][]{2012A&A...548A..10M,2014A&A...564A..27D}. Models of red giant core angular momentum transport including coupling from the Taylor-Spruit dynamo predict WD rotation rates an order of magnitude faster than observations imply \citep[e.g.,][]{2014ApJ...788...93C} and that additional viscosity after the core-He burning phase is required to explain the WD rotation rates \citep[e.g.,][]{2019A&A...622A.187D}. A modified formulation of the Taylor instability has been claimed to provided the additional torque needed to brake the stellar core rotation \citep{2019MNRAS.485.3661F}, though this formalism has been called into question \citep{2020A&A...634L..16D}.  \citet{2018ApJ...868..150T} combine surface and core rotation rate measurements for core-He burning stars to show that core-He burning stars are also rotating slower than predicted; they conclude that both enhanced mass loss and radial differential rotation in the surface convection zone may be needed to explain the observations. In short, theoretical models for angular evolution transport in evolved stars do not satisfactorily reproduce the observed core rotation rates.  

Observed WD rotational periods useful for the study of angular momentum evolution have been derived primarily via asteroseismology, especially through space-based observations made by \emph{Kepler}, K2, and TESS.  \citet{2017ApJS..232...23H} greatly increased the number of WDs with asteroseismic rotational periods to confirm that WDs with $M\lesssim 0.7\,\msun$ generally rotate at periods $\gtrsim 5$ hr. These observations have proven crucial for constraining models of angular momentum transport in the evolved stars discussed above.

Similarly, one would expect that the rotational periods of higher mass WDs, i.e., WDs originating from stars with initial masses $> 3\,\msun$, would be particularly useful for constraining angular momentum transport in the evolution of intermediate-mass stars.  However, very few observational constraints from massive WDs are available.  A primary reason for this is the steepness of the initial mass function, so such massive WDs are relatively rare.  \citet{2017ApJS..232...23H} report asteroseismic rotation measurements for only three WDs with $M_\mathrm{WD} > 0.72\,\msun$ (progenitor masses $> 3\,\msun$), though interestingly all three WDs have rotation periods $<10$ h.  In particular, \citet{2017ApJ...841L...2H} present the fastest known rotation period derived from asteroseismology, the 1.13 h rotation of the 0.87 \msun WD \object{EPIC 211914185}.  

A growing number of isolated  massive WDs have rotational periods determined via photometric variability. Many of these are very rapid rotators, especially in comparison to WDs with typical masses of $\approx 0.6\,\msun$.  Examples include (in order of increasing period):  the 70 s rotation of the 1.27\,\msun \object[SDSS J221141.80+113604.3]{SDSS J2211$+$1136} \citep{2021ApJ...923L...6K}, the 353 s rotation of the 1.33\,\msun \object[Gaia DR2 4479342339285057408]{WD J$1832+0856$}  \citep{2020MNRAS.499L..21P}, the 416 s rotation of the $1.327 - 1.365$ \msun \object{ZTF J$1901+1458$} \citep{2021Natur.595...39C}, the 725 s rotation of the 1.34 \msun \object[WD 0316-849]{EUVE J0317-855} \citep{1997MNRAS.292..205F}\footnote{This rotator is not technically isolated, as it has a common proper motion companion WD with $\approx 7\arcsec$ separation, but the projected separation is $\approx 200$ AU \citep{2010A&A...524A..36K}. Therefore, accretion powered spin-up is unlikely to be the cause of rapid rotation. However, a relatively large number of widely separated WD+WD appear to have been triples that have had an inner pair merge \citep{2022arXiv220600025H}.}, and the 2289 s period of the 0.97 \msun \object[SDSS J152934.98+292801.9]{SDSS J1529+2928} \citep{2015ApJ...814L..31K,2020ApJ...898...84K}. \citet{2021ApJ...912..125G} announced the likely rotation of \object{ZTF J0534+7707} with $P=2600$ s, though the mass ($0.7 - 0.8$ \msun) is not yet well constrained.  \citet{2013ApJ...773...47B} present a compilation of magnetic WDs with measured rotation periods; their massive WDs have a range of periods from $\approx 2 - 100$ hr.
\object{EPIC 228939929} has a claimed rotation period of  317 s, but the WD mass is poorly constrained and likely $\lesssim 0.6$ \msun \citep{2017ASPC..509....3D,2020ApJ...894...19R,2020ApJ...898...84K}, and recent observations suggest the true rotation period may be a more leisurely 635 s (J. Farihi, private communication).  

The application of photometrically derived rotation periods to evolutionary models of intermediate mass stars is complicated by multiple factors.  First, the presence of the magnetic fields required to create a spotted WD influences the rotational evolution of the stellar core during post-main sequence evolution \citep[e.g.,][]{2004IAUS..215..561K}. Second, a significant fraction of massive WDs likely formed via merger events \citep[e.g.][]{2020A&A...636A..31T,2020ApJ...891..160C}, which may also be the origin of strong magnetic fields in WDs \citep[e.g.,][]{2000PASP..112..873W,2008MNRAS.387..897T,2011PNAS..108.3135N,2012ApJ...749...25G}.  Although extraordinarily interesting objects in their own rights, the rotation of merger remnants does not directly illuminate the problem of angular momentum evolution in single stars.  Third, the true cause of observed photometric variability can require substantial effort to confirm \citep[e.g.,][]{2009A&A...496..813M}, be it rotation, ellipsoidal variations, irradiation or reflection from a close binary, grazing eclipses, etc. 

These concerns are significant and should be forefront in the minds of those interpreting the rotational periods of massive WDs.  Even so, it is desirable to increase the number of massive WDs with measured rotational periods in case the distribution of rotational periods reveals features that can identify the origin of a WD's rotation, be it the result of a merger or of single star evolution.

In this paper, we identify and quantify previously unreported photometric variability in the massive WD EGGR 156.
EGGR 156 is a DA WD whose spectrum exhibits obvious Zeeman splitting corresponding to a mean magnetic field strength of $\approx 16.1$ MG \citep{2009A&A...506.1341K}.  Parameter fits derived from SDSS $u$, Pan-STAARS $grizy$, and Gaia DR2 precision parallaxes give $T_\mathrm{eff}=13410\pm 130$ K and $M=1.298\pm 0.003\, \msun$ assuming a pure H atmosphere \citep{2020ApJ...898...84K}.  In Section \ref{sec.photobs} we present details of the photometric observations and detailed analysis of the photometric variability.  We then discuss the likely nature of the variability in Section \ref{sec.nature}, the likely origin of the WD in Section \ref{sec.merger}, and the potential to detect period changes in rapidly rotating magnetic stars in Section \ref{sec.spindown}.

\section{Photometric Observations} \label{sec.photobs}
We obtained high-speed photometric observations of EGGR 156 as part of a pilot program to search for rotation in massive, magnetic WDs. We selected the target by searching the Montreal White Dwarf Database \citep{2017ASPC..509....3D} for magnetic DA WDs more massive than 1.0 \msun with no known variability and well-placed to observe from McDonald Observatory during an RA gap in our primary program. Once we had detected variability, we prioritized the collection of additional data.   We collected time-series photometry over multiple nights during July and November 2020.  A summary observing log is in Table \ref{tab.obslog}.   

\begin{deluxetable}{clccccccDl}
\tablewidth{0pt}
\tabletypesize{\scriptsize}
\tablecaption{Observing Log for High-Speed Photometry \label{tab.obslog}}
\tablehead{\colhead{Series} & \colhead{UT Date} & Filter & \colhead{Midpoint of} & \colhead{Exp} & \colhead{Series} & \multicolumn2c{FWHM} & \multicolumn2c{Airmass} & \colhead{Remarks} \\ & & & 1\textsuperscript{st} Exposure & Time & Length & Mean & Std. Dev. & \multicolumn2c{Coeff.} & \\ & & & (UTC) & (s) & (s) &  ($\arcsec$) & ($\arcsec$) & &  & }
\startdata
\decimals
1 & 2020 Jul 25 & BG40 & 07:44:09.0 & 10 & 12,480 & 1.03 & 0.10 & $0.0005$ & Highly variable cloud extinction \\
2 & 2020 Jul 28 & BG40 & 06:25:59.0 & 10 & 16,880 & 1.63 & 0.33 & $-0.0086$ & \\
3 & 2020 Jul 29 & BG40 & 07:21:45.0 & 10 &   7220 & 1.57 & 0.09 & $-0.0075$ & \\
4 & 2020 Nov 11 & $g$  & 01:50:54.5 & 15 & 15,981 & 1.70 & 0.37 & $-0.0070$ & \\
  &             & $r$  & 01:51:41.5 & 15 &        & 1.62 & 0.43 & $-0.0005$ & \\
  &             & $i$  & 01:52:28.5 & 15 &        & 1.60 & 0.45 & $0.0018$ & \\
5 & 2020 Nov 12 & BG40 & 00:58:21.0 & 10 &   6150 & 1.08 & 0.12 &  $-0.0078$ & Thickening clouds after 5150 s 
\enddata
\end{deluxetable}

We obtained all time-series observations with the ProEm frame-transfer CCD on the McDonald Observatory 2.1 m Otto Struve Telescope at Cassegrain focus. For most observations we used a BG40 Schott glass filter with 10 s exposures. On UT 2020 November 11, we cycled continuously between SDSS $g$, $r$, and $i$ filters, taking three consecutive 15 s exposures in each filter, followed by a 2 s gap as we rotated the filter wheel.  We collected bias, dark frame, and dome flat field images each night prior to sunset.

We reduced data and extracted relative aperture photometry using AstroImageJ v.\ 4.0.0.1 \citep{2017AJ....153...77C}. We utilized a variable aperture scaled with the full-width half-maximum (FWHM) of the target such that the aperture radius $r=1.4\times\mathrm{FWHM}$.  The only useful comparison star in the ProEM field of view is at $\mathrm{RA}=$ 22:57:30.13, $\mathrm{Dec}=+$07:56:36.2 \citep[J2000; coordinates from SDSS DR16,][]{2020ApJS..249....3A}; this star is in the Gaia EDR3 archive with ID 2712263571022741248, $G=14.916$, and $G_\mathrm{BP}-G_\mathrm{RP}=1.158$ \citep{2016A&A...595A...1G,2021A&A...649A...1G,2021A&A...649A...3R}.  

\begin{figure*}
    \begin{minipage}{0.49\columnwidth}
    \centering
    \includegraphics[clip, trim=1cm 5cm 1cm 3cm, width=\textwidth]{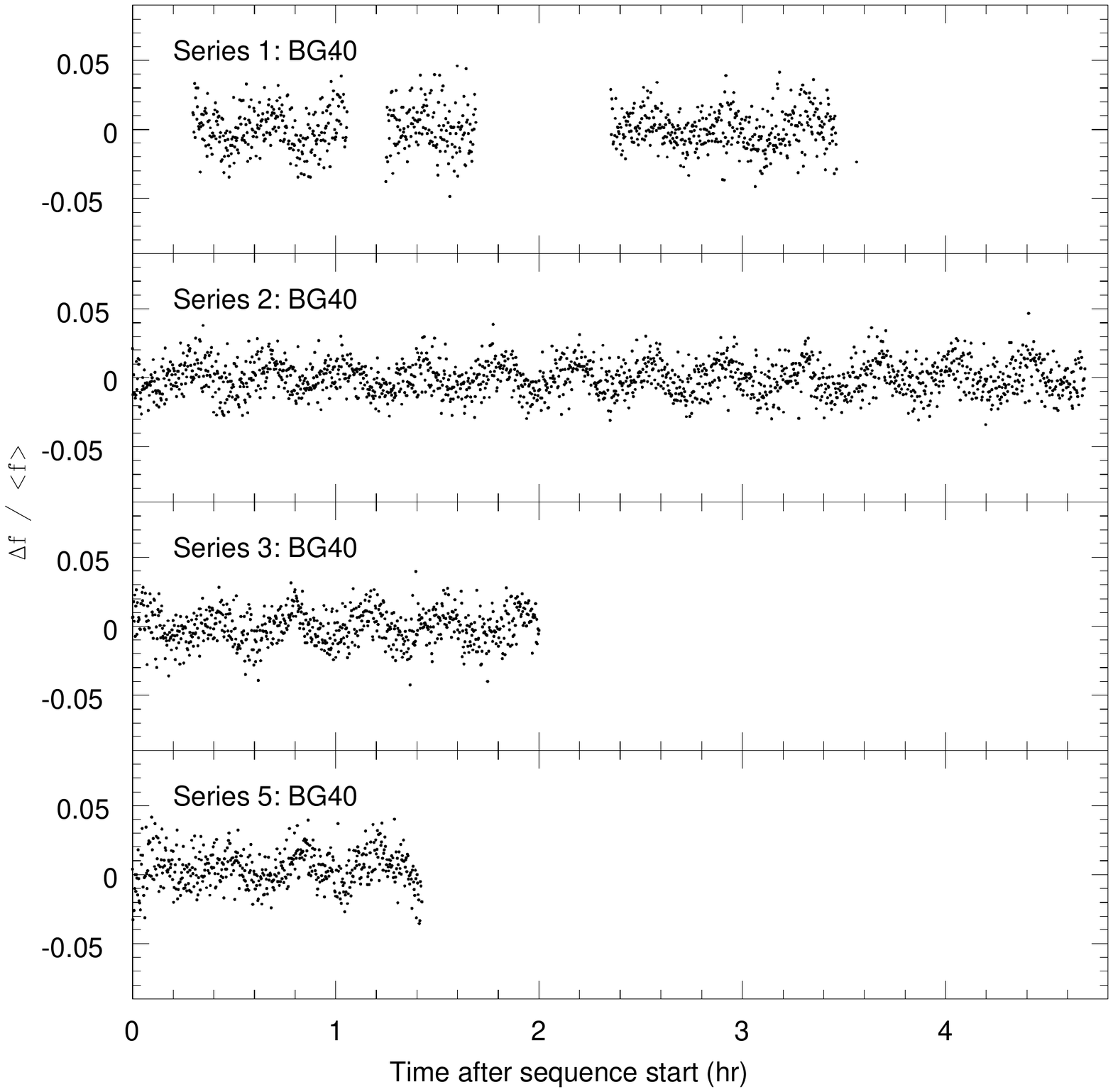}
    \end{minipage}
    \hfill

    \begin{minipage}{0.49\columnwidth}
    \centering
    \includegraphics[clip, trim=1cm 5cm 1cm 5cm, width=\textwidth]{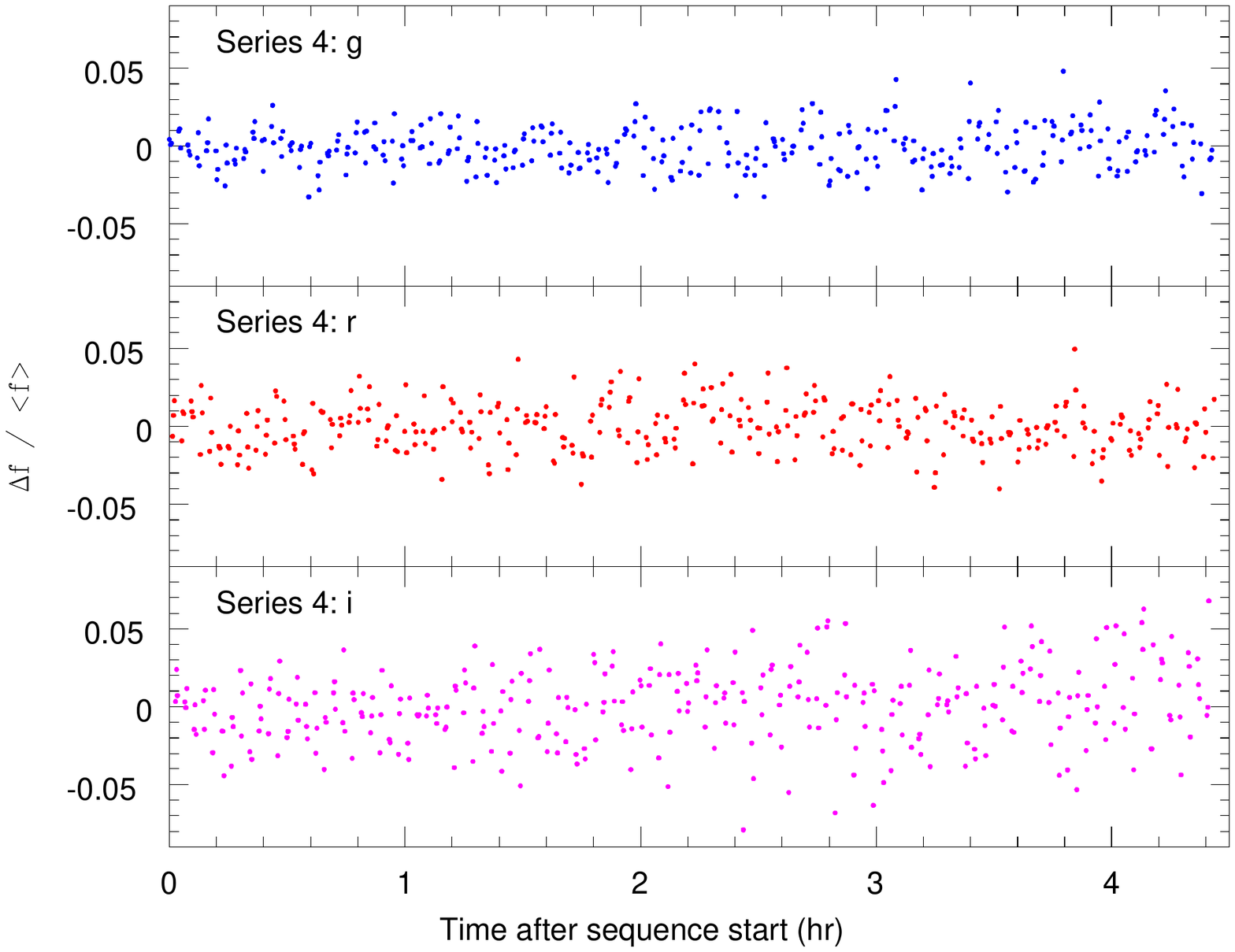}
    \end{minipage}
    \caption{Light curves for all time series observations of EGGR 156.  \emph{Left:} All BG40 time series observations. \emph{Right:} The interleaved $gri$ observations.  In all panels, the horizontal axes give the time in hours from the start of each series (see Table \ref{tab.obslog}). The vertical axes show the fractional change in flux compared to the mean flux for that filter.  The modulation can be seen by visual inspection in the BG40 band (left panels) and the $g$ band (top right panel), but is barely visible in $r$ (middle right panel) and not at all in $i$ (bottom right panel).  }
    \label{fig.lcs}
\end{figure*}

We fit and divided the extracted relative light curves by a linear function of airmass.  We corrected times of individual exposures to Barycentric Dynamical Time using tools in AstroImageJ.  The FWHM and detrending coefficients for each observational series are given in Table \ref{tab.obslog}.  

Visual inspection of the resulting time series data reveals obvious periodic variability with a period of $\sim 20$ min.  We present the normalized relative flux light curves for all observing series in Figure \ref{fig.lcs}.

We performed our frequency analysis using Period04 \citep{2005CoAst.146...53L}.  We excluded portions of Series 1 and Series 5 BG40 data that were heavily extincted by clouds.  We also clipped any extreme outlying data points, usually caused by cosmic ray hits.  Errors on the frequencies, amplitudes, and phases of all modulations were determined both by least-squares and Monte Carlo methods within Period04; in all cases, the Monte Carlo uncertainties were larger, and so we adopt these values.

\begin{figure}
    \centering
    \includegraphics[clip, trim=1cm 5cm 1cm 3cm, width=0.9\columnwidth]{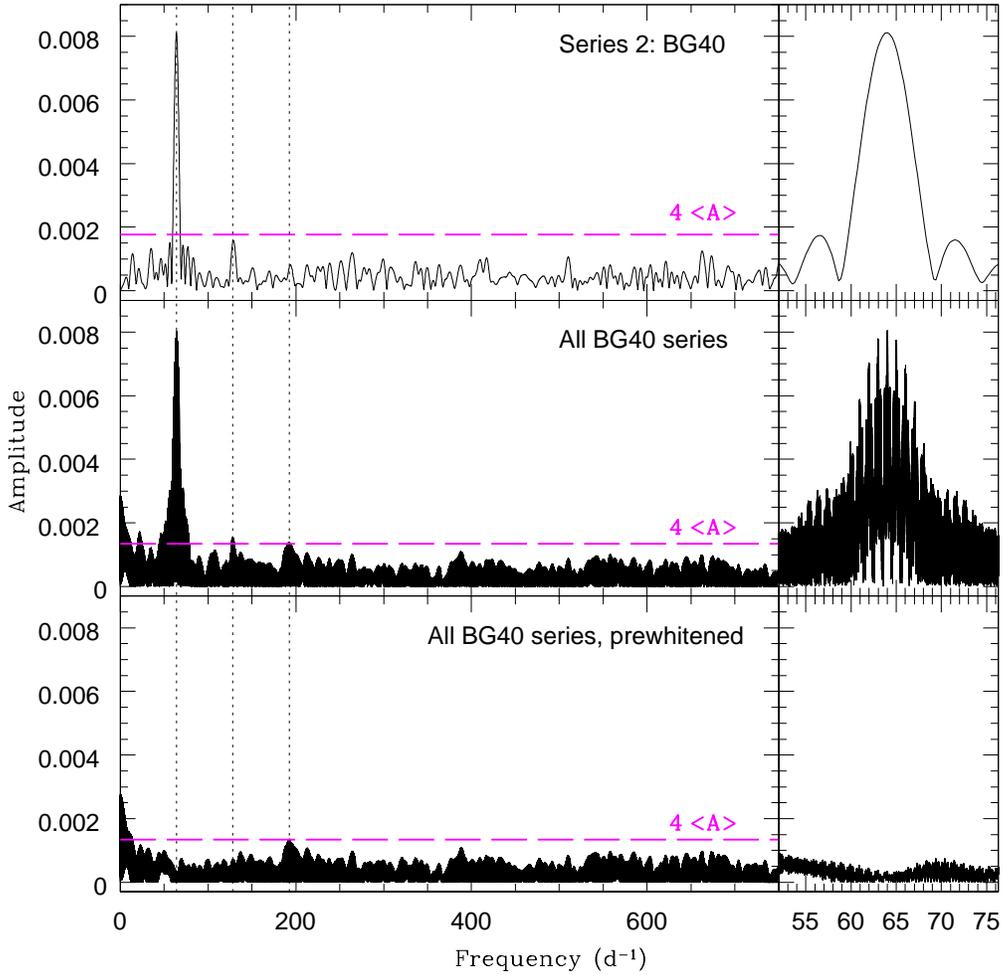}
    \caption{Periodograms for EGGR 156 (left panels) and the spectral window function for the fundamental frequency (right panels) scaled to the fit amplitude.  Shown are the periodograms for Series 2 (top panel) and the entire combined BG40 data set (middle panel).  Horizontal long-dashed, magenta lines show the adopted $4\avgamp$ significance threshold.  The bottom panel shows the entire data set after prewhitening by the fundamental frequency $f_0$ and the harmonic $2f_0$; no residual significant peaks remain.  Vertical dotted lines indicate the location of harmonics $f_0$, $2f_0$, and $3f_0$.}
    \label{fig.bg40_dfts}
\end{figure}

We began our analysis with the data from Series 2, as that series was the longest and of generally excellent quality.  The periodogram from this series is plotted in the top panel of Figure \ref{fig.bg40_dfts}. We identified potentially significant peaks in the periodogram as those peaks with amplitudes larger than $4\avgamp$, where $\avgamp$ is the average amplitude of all sampled frequencies between 0 d$^{-1}$ and 4320 d$^{-1}$, the Nyquist frequency.  This criterion has been found to exclude  $\approx 99.9\%$ of spurious peaks \citep{1993A&A...271..482B,1997A&A...328..544K}. For series 2 data, $4\avgamp = 0.177\%$ over 844 sampled frequencies; for the combined BG40 data set, $4\avgamp = 0.134\%$ over 2136 sampled frequencies.  

A highly significant signal is observed at a frequency of $f_0=64.15\pm 0.12$ d$^{-1}$, corresponding to a period of $22.45\pm 0.04$ m with an amplitude of $0.81\%\pm 0.04\%$  ($18.4\avgamp$ for Series 2).  A likely harmonic is also observed at $2f_0$ at an amplitude of $0.16\pm 0.04\%$, though this amplitude is just below our significance threshold.

\begin{deluxetable}{lCCCCCCC}
\tablewidth{0pt}
\tablecaption{Sinusoidal Fit Parameters \label{tab.freqs}}
\tablehead{\colhead{Data} & f_0 & P_0 & A_0 & \phi_0 \tablenotemark{a} & f_1\tablenotemark{b} & A_1 & \phi_1 \tablenotemark{a} \\
 & \mathrm{d}^{-1} & \mathrm{minutes} & \% & & \mathrm{d}^{-1} & \% & \\}
\startdata
Series 2 (BG40) & 64.15\pm 0.12 & 22.45\pm 0.04 & 0.815\pm 0.037 & 0.412\pm 0.007 & 128.3 & 0.157\pm 0.036 & 0.436\pm 0.036 \\
All BG40 & 64.15039\pm 0.00016 & 22.44725\pm 0.00006 & 0.806\pm 0.027 & 0.4115\pm 0.0053 & 128.30077 & 0.141\pm 0.027 & 0.387\pm 0.031 \\
Series 4: $g$ & 64.15039\tablenotemark{c} & 22.44725\tablenotemark{c} & 0.863\pm 0.093 & 0.421\pm 0.017 & \nodata & \nodata & \nodata \\
Series 4: $r$ & 64.15039\tablenotemark{c} & 22.44725\tablenotemark{c} & 0.68\pm 0.11 & 0.484\pm 0.027 & \nodata & \nodata & \nodata \\
Series 4: $i$ & 64.15039\tablenotemark{c} & 22.44725\tablenotemark{c} & 0.18\pm 0.19 & 0.70\pm 0.16 & \nodata & \nodata & \nodata \\
\enddata
\tablenotetext{a}{ Phase at $t_0 = \mathrm{BJD}\, 2459055$ (exact) }
\tablenotetext{b}{Frequency constrained to be = $2f_0$}
\tablenotetext{c}{Frequency constrained to result from combined BG40 data}
\end{deluxetable}

We then improved the frequency fit by adding iteratively each individual data series and redetermining the best-fit frequency, amplitude and phase of the modulation, first by adding Series 3, then Series 1, and finally Series 5 data.  In each iteration, only a single frequency alias was consistent with the best-fit frequency and uncertainty from the previous iteration.  We provide the resulting fit sinusoidal parameters in Table \ref{tab.freqs}.  

\begin{figure}
    \centering
    \includegraphics[clip, trim=1cm 5cm 1cm 3cm, width=0.9\columnwidth]{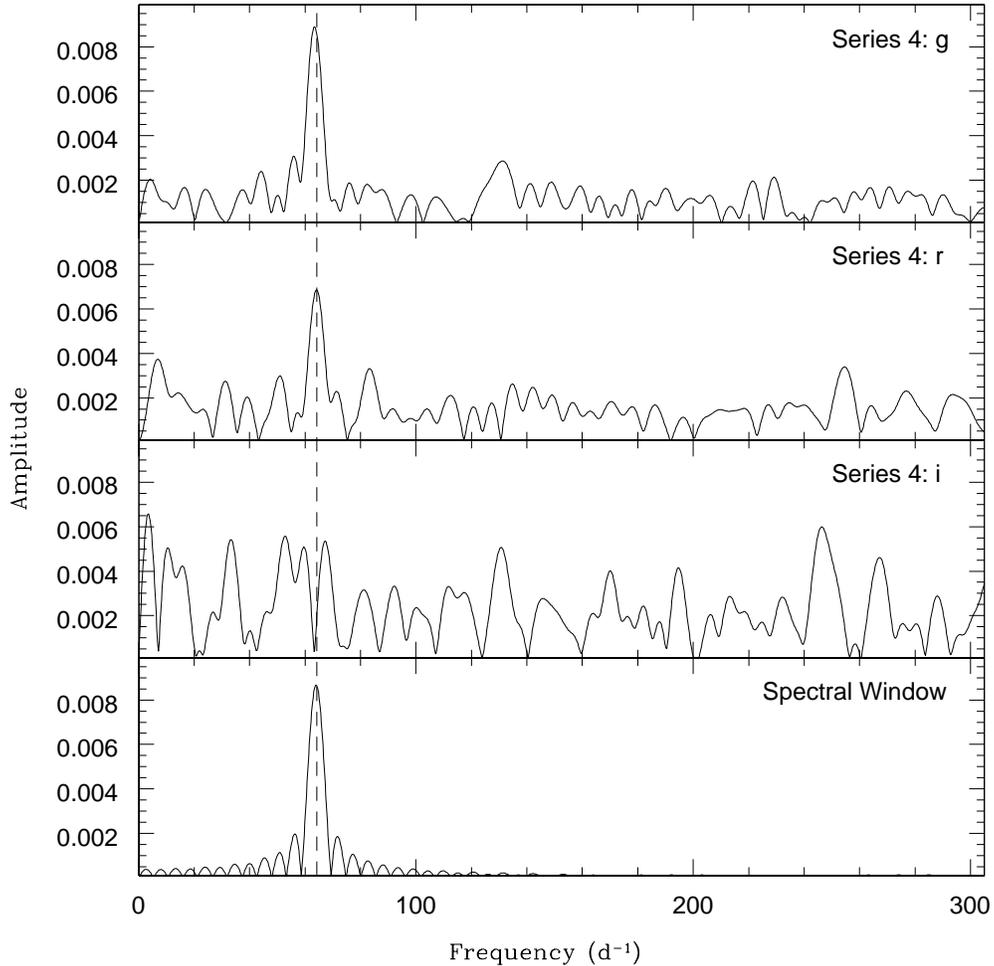}
    \caption{Periodograms for the $g$ (top panel), $r$ (second panel), and $i$ (third panel) observations of EGGR 156.  The bottom panel shows the spectral window for the frequency $f_0$ scaled to the $g$-band fit amplitude.  No significant $i$-band signal is observed, although the noise is significantly higher in that band.  The vertical dashed line indicates the best frequency as determined from BG40 observations.}
    \label{fig.gri_dfts}
\end{figure}

For the multi-filter observations of Series 4, we fixed the frequency of the modulation to that from the combined BG40 dataset, and we then fit the amplitudes and phases for each filter separately.  We plot the individual periodograms for each filter in Figure \ref{fig.gri_dfts}, and provide the best-fit sinusoidal parameters in Table \ref{tab.freqs}.

From the parameters in Table \ref{tab.freqs} we observe two trends.  First, the amplitude of the variability is higher at shorter wavelengths, with a $g/r$ amplitude ratio of $\approx 1.3$, strikingly similar to that observed in rotating strongly magnetic WDs of similar temperatures \citep[e.g.,][]{2018RNAAS...2...27S}, though amplitude ratios in pulsation modes of ZZ Ceti stars are also similar \citep[e.g., ][]{1982ApJ...259..219R,1995ApJ...438..908R}.

We also find marginal evidence of a phase lag in longer wavelength filters. The $r$-band maximum lags the $g$-band maximum by $0.063\pm 0.032$ cycles ($\approx 85$ s), and the $i-$ band maximum lags the $g$-band by a phase difference of $0.28\pm 0.16$, if the sinusoidal fit is to be believed given the low amplitude and significant noise in the $i$-band periodogram (Fig.~\ref{fig.gri_dfts}).  The BG40 phase is fully consistent with the $g$-band phase, which is expected given that its bandpass overlaps the $g$ filter but only a portion of $r$. Wavelength-dependent phase lags are observed in other rotating magnetic WDs \citep[e.g.,][]{1997MNRAS.292..205F,2020ApJ...894...19R,2021Natur.595...39C} but are not observed in studies of pulsating WDs \citep[e.g., ][]{1995ApJ...438..908R}.

\section{Discussion}
\subsection{Nature of the Variability}
\label{sec.nature}
\subsubsection{Rapid Rotation of a Spotted Star?}
\label{sec.nature.var}

The characteristics of the photometric variability in EGGR 156 -- a single period of modulation (with harmonics), a period stable over a baseline of months, and a potential phase lag as a function of wavelength -- are consistent with the interpretation that its variability is caused by rapid rotation of WD with a magnetic spot.

\begin{figure}
    \centering
    \begin{minipage}{0.65\textwidth}
    \centering
    \includegraphics[clip, trim=1cm 5cm 1cm 3cm, width=\linewidth]{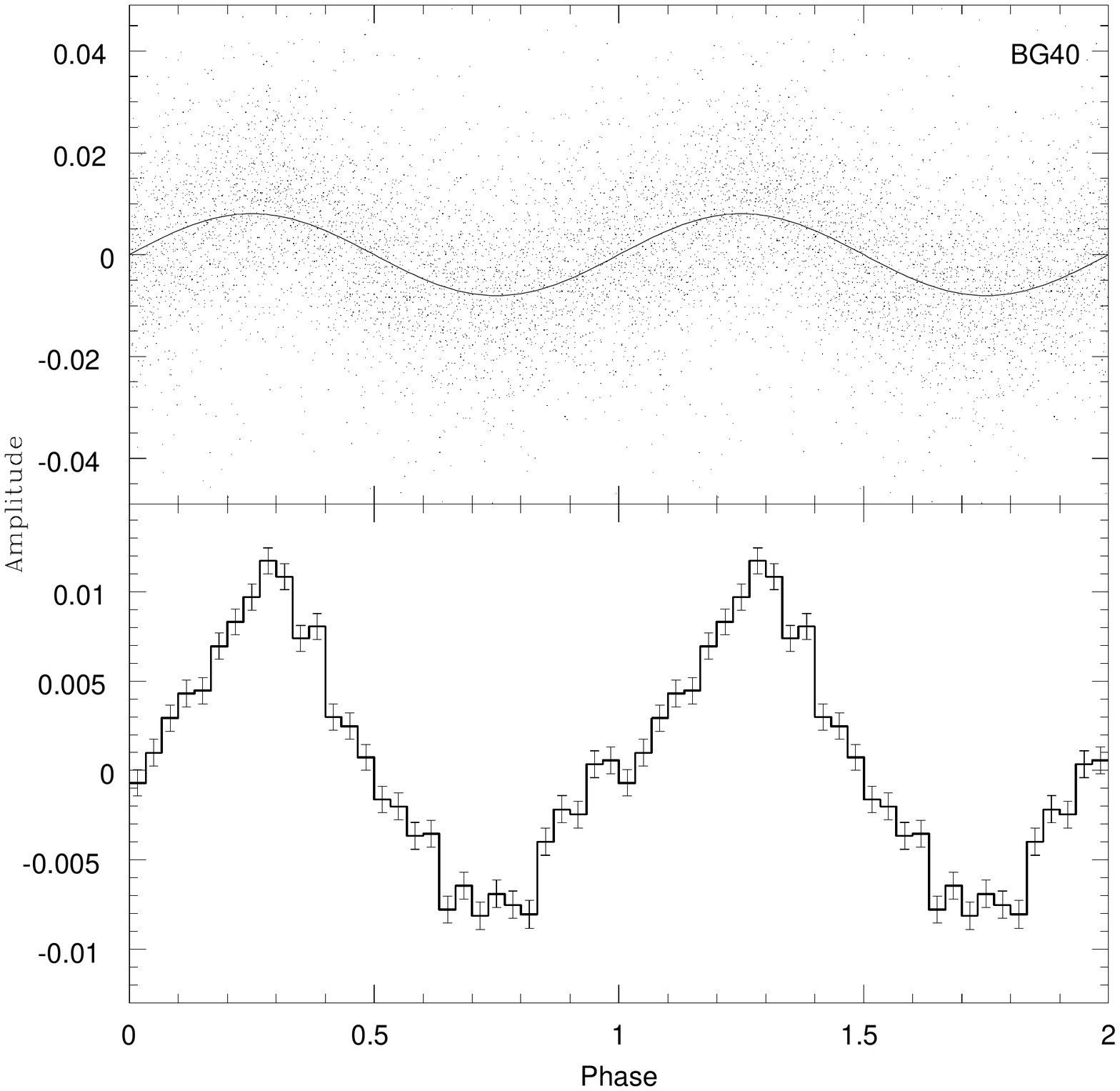}
    \end{minipage}%
    \begin{minipage}{0.33\textwidth}
        \includegraphics[clip, trim=6cm 5cm 7cm 3cm, width=0.9\linewidth]{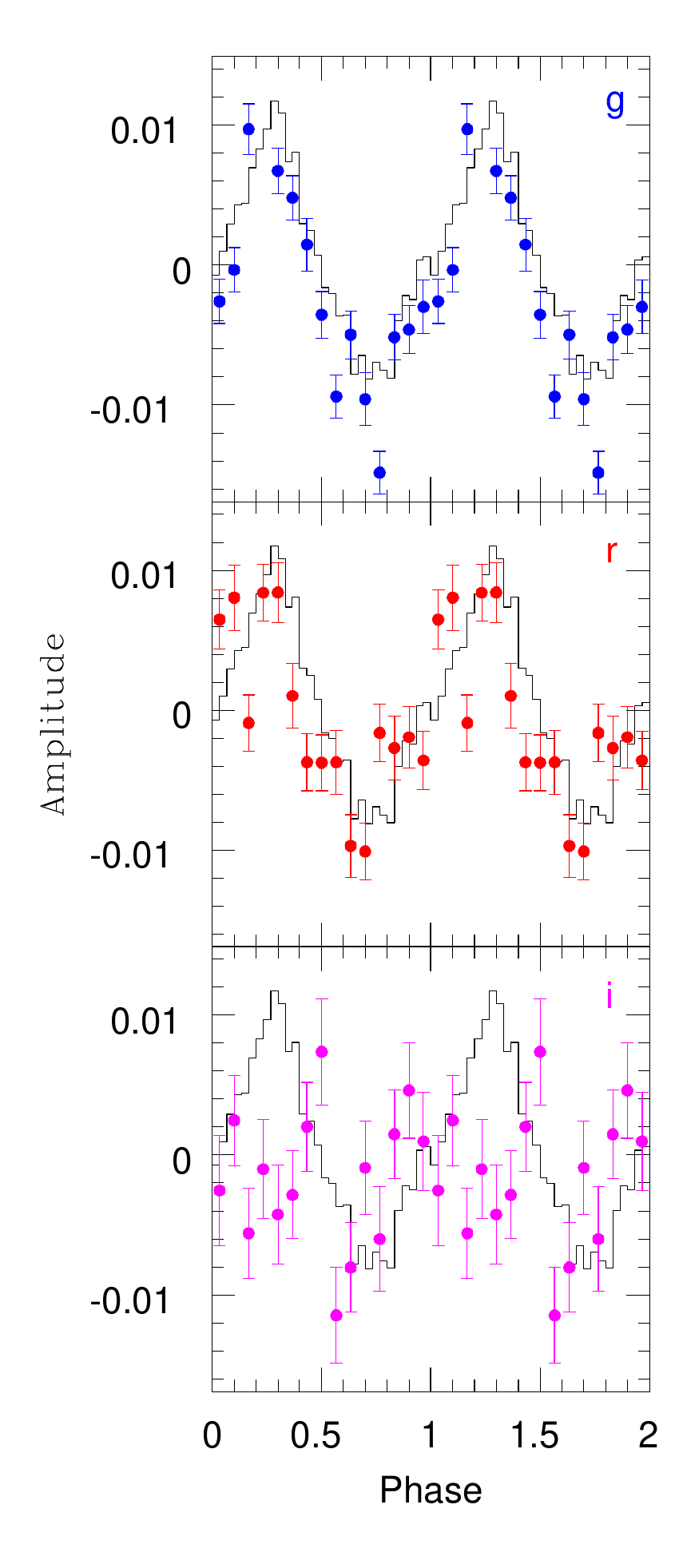}
    \end{minipage}
    \caption{\emph{Left:} All BG40 data folded at best-fitting frequency (top) and binned into 30 phase bins (bottom).  The solid line in the top plot is the best-fitting sinusoid.  Error bars on the bottom panel indicate the uncertainty in the mean for that bin.  \emph{Right:} Points with error bars are phase-folded and binned lightcurves for $g$ (top), $r$ (middle), and $i$ (bottom) observations compared to the BG40 binned light curve (histogram).  The $gri$ data were binned into 20 bins; again the error bars show the uncertainty in the mean for the bin. The folded light curves in $r$ and $i$ show qualitative evidence of phase lag.}
    \label{fig.lc_folded}
\end{figure}

Figure \ref{fig.lc_folded} shows the light curve folded upon the 22.44725 min period derived using the entire BG40 data set for both unbinned BG40 data (top left panel) and binned data in BG40 (bottom left panel).  The luminosity maxima are slightly peaked compared to a sinusoid, and the luminosity minima are slightly broadened.  We observe no evidence of eclipses.  

Further, as can be seen in the folded multi-band photometry (right panels of Figure \ref{fig.lc_folded}), the photometric modulations are larger in amplitude at shorter wavelengths.  This likely rules out a close binary with an irradiated, cooler WD companion, as irradiation should show higher amplitudes at longer wavelengths \citep{2021A&A...647A.184R}.

We also examined the individual SDSS spectral exposures for EGGR 156 to look for radial velocity variations that might be expected from a close binary system.  The spectra were taken consecutively with exposure times of 900 s, 1200 s, and 1500 s.   We observed no significant radial velocity differences, but this is perhaps unsurprising.  Each exposure integrates over a significant fraction of the variability period of EGGR 156, and thereby should nearly completely average over any spectral variations. 

As part of an ongoing project described in \citet{2020IAUS..357...75H}, the SDSS spectrum of EGGR 156 has recently been re-fit with a magnetic hydrogen atmosphere model (P.~Dufour, private communication).  While the resulting photometric model is consistent with the parameters published by \citet{2020ApJ...898...84K} and the magnetic field strength derived by \citet{2009A&A...506.1341K}, the observed Balmer lines are significantly weaker than predicted by the atmospheric models.  This discrepancy could be explained if EGGR 156 were an unresolved binary, but the allowable parameter space for such a binary is extraordinarily small.

Another possible explanation for the poor Balmer line fits could be due to a complex magnetic field geometry changing as a function of rotational phase.  We again examined the individual SDSS spectral exposures to explore whether the magnetic features changed significantly between exposures, potentially indicative of a complex magnetic field geometry, which might explain the poor spectral fit. Again we observed no qualitative differences, and again this is not surprising due to the large phase covered by each spectroscopic integration.  

\subsubsection{A Highly Magnetic ZZ Ceti star?}\label{sec.nature.zzceti}
Although we strongly suspect that the photometric variability of EGGR 156 is caused by rapid rotation of a WD with a magnetic spot, we also mention the possibility that EGGR 156 might be a ZZ Ceti star exhibiting a single dominant mode of pulsation.  

\citet{2017MNRAS.468..239C} present the discovery of four massive ZZ Ceti WDs, three of which have only a single significant mode of pulsation, though \citet{2021ApJ...923L...6K} suggest that these three WDs may actually be rotating instead of pulsating.  EGGR 156 is located within the ZZ Ceti instability strip presented by  \citet{2011ApJ...743..138G}, and likewise exhibits a single mode of variability.  Further, pulsation amplitudes are higher at shorter wavelengths  \citep[e.g.][]{1982ApJ...259..219R}, as we observe for EGGR 156.

However, there are no unambiguously pulsating WDs known with magnetic fields larger than a few kilogauss \citep[e.g.,][]{1994ApJ...430..839W,2008ApJ...683L.167D,2019A&ARv..27....7C}.  \citet{2008ApJ...683L.167D} report that the hot DQ WD \object[SDSS J142625.71+575218.3]{SDSS J1426+5752} may be a strongly magnetic pulsator, but evidence from other hot DQVs suggests its variability is due to rotation instead \citep{2016ApJ...817...27W}.  In addition, the presence of strong magnetic fields should suppress the convective energy transport that drives pulsations \citep[e.g.,][]{2015ApJ...812...19T}.  Pulsations should also show coherence in phase as a function of wavelength, while EGGR 156 has marginal detection of wavelength-dependent phase lags. In the absence of clear signatures of non-radial pulsations, such as the presence of multiple significant independent modes of variability from EGGR 156, rotation is the more conservative interpretation for the observed photometric variations.

\subsection{Merger Remnant or Evolved Intermediate Mass Star?} \label{sec.merger}

Assuming that EGGR 156 is rapidly rotating, we are still left with the problem of ascertaining the ultimate source of the WD's high angular momentum.  The two most plausible sources would be either residual angular momentum from the merger of two lower-mass WDs, or primordial angular momentum from the rapid rotation of an intermediate mass progenitor star.  

Although there is no definitive evidence for either scenario, there are some observations that support the merger origin for EGGR 156.   The WD mass of 1.3 \msun is very close to twice the mean mass of the sharply peaked field WD mass distribution \citep[$\left < M \right > =0.601\,\msun$;][]{2019ApJ...876...67B}, as might be expected for a random pairing of WDs.  Second, recent evolutionary models for WD mergers suggest that typical remnants should have rotation rates of $\approx 10 - 20$ m \citep{2021ApJ...906...53S}, similar to our observed 22.4 m period.  Third, the transverse velocity of EGGR 156 as calculated from its Gaia EDR3 parallax and proper motion \citep[$\varpi=21.97\pm 0.10$ mas; $\mu = 235.78\pm 0.12$ mas yr$^{-1}$; ][]{2021A&A...649A...1G} is 50.9 km s$^{-1}$, which is indicative of a merger origin for a massive field WD \citep{2015ASPC..493..547D,2020ApJ...898...84K}. Fourth, the parallax, kinematics and photometry of EGGR 156 place it close to the Q branch on the Hertzsprung-Russell diagram, which have been posited to be ultra-massive WD merger remnants \citep{2019ApJ...886..100C,2020ApJ...891..160C}.

If EGGR 156 is the remnant of the single star evolution of an intermediate mass star, then we can estimate the progenitor star mass using an extrapolation of the initial-final mass relation (IFMR). Using the IFMR of \citet{2009ApJ...693..355W}, we calculate $M_\mathrm{init} = 7.4\,\msun$, and from the IFMR of \citet{2018ApJ...866...21C}, we calculate $M_\mathrm{init}=7.7 - 8.8\,\msun$, depending on the selected stellar evolutionary model.  In either case, we would be looking at the oxygen-neon remnant of a star that did not experience a He flash.  

As discussed by \citet{2015ASPC..493...65K}, such white dwarfs could have rotation periods as short as 400 s in the absence of strong core-envelope coupling during the giant phases of evolution.  The fact that EGGR 156 is rotating at rates approaching this theoretical maximum in spite of having a 16 MG magnetic field would imply that even moderately large primordial magnetic fields have only a minimal impact on angular momentum evolution of the AGB core.

\subsection{The Potential to Detect Spin-down}\label{sec.spindown}

Given the rapid rotation of this highly magnetized star, we explore the potential that long-term monitoring may be able to detect spin-down of the star due to magnetic dipole radiation, the same spin-down mechanism assumed for pulsars.  
Nearly a half-century of time series observations of the pulsating WD \object{G117-B15A} constrain its period change, $\dot{P}$, to a precision of $\approx 10^{-15}$ s s$^{-1}$ \citep{2021ApJ...906....7K}.  A precision of $\approx 5\times 10^{-14}$ s s$^{-1}$ has been reported in detection of the spin-down rate of the WD pulsar \object{AR Sco} over a time baseline of 7 yr \citep{2018AJ....156..150S}.
If the $\dot{P}$ of EGGR 156 is of a similar order of magnitude, the spin-down may be detectable within an astronomer's career.

The spin-down of a rotating uniformly magnetized sphere due to magnetic dipole radiation is given by:
\begin{equation}
    \dot{P} = \frac{8\pi^2}{3c^3} \frac{\sin^2{\alpha}\, R^6 B^2}{IP_{\rm rot}}
\end{equation}
in $cgs$ units, where $R$ is the stellar radius, $B$ is the surface magnetic field, $\alpha$ is the inclination of the magnetic field to the rotational axis, $I$ is the moment of inertia, and $P_{\rm rot}$ is the rotation period \citep{2016era..book.....C}.  

If we assume the WD is a uniform density sphere and adopt $\sin^2{\alpha} =\, <\!\sin^2{\alpha}\!>\, = 1/2$, the equation becomes 
\begin{equation}
    \dot{P} = \frac{10 \pi^2}{3 c^3} \frac{R^4 B^2}{M P_{\rm rot}}
\end{equation}
where $M$ is the mass of the WD.  We calculate $R=3.0\times 10^8$ cm from the mass and surface gravity of EGGR 156 reported in \citet{2020ApJ...898...84K} and assume a uniform magnetic field $B=16.1$ MG to determine a spin-down rate of $\dot{P} \approx 7.4\times 10^{-19}$ s s$^{-1}$, or roughly 3 orders of magnitude smaller than the timing precision achieved by \citet{2021ApJ...906....7K}.  A similar calculation for the 70 s period rotator SDSS J2211+1136 \citep{2021ApJ...923L...6K} finds a larger but still undetectable spin-down rate of $\dot{P} \approx 3.7\times 10^{-18}$ s s$^{-1}$.

Since $\dot{P}$ scales by $B^2$, WDs with significantly stronger magnetic fields may have more easily detectable $\dot{P}$.  We therefore repeat the above calculation for two highly magnetic, rapidly rotating, ultramassive WDs:  ZTF J190132.9+145808.7 and RE J0317$-$853. For ZTF J190132.9+145808.7 we use the average parameters reported by \citet{2021Natur.595...39C}: $M=1.34\,\msun$, $R=2.14\times 10^8$ cm, $P_{\rm rot}=416.2$ s, and $B\approx 800$ MG.  The resulting magnetic dipole spin-down rate is $\dot{P}\approx 1.5\times 10^{-15}$ s s$^{-1}$, which would be on the edge of detectability given a data set similar to that used by \citet{2021ApJ...906....7K}.  

For RE J0317$-$853, we adopt mid-range values of physical parameters from \citet{2010A&A...524A..36K}: $M=1.33\,\msun$, $R=0.00353\, R_\odot$, and from \citet{1997MNRAS.292..205F} we adopt $P_{\rm rot}=725.7377$ s and $B=317.5$ MG.  The result is $\dot{P} \approx 
2.3\times 10^{-16}$ s s$^{-1}$, just below current detection limits.

In short, it is reasonable to consider the potential for constraining physical details of rapidly rotating, ultramassive, very highly magnetic WDs via magnetic dipole radiation through the use of long-term, precision time-series observations.  However, most of the known rapidly rotating massive magnetic WDs should be stable and precise clocks useful for experiments searching for other sources of torques or timing changes.

\begin{acknowledgements}
This work has made use of data from the European Space Agency (ESA) mission {\it Gaia} (\url{https://www.cosmos.esa.int/gaia}), processed by the {\it Gaia} Data Processing and Analysis Consortium (DPAC, \url{https://www.cosmos.esa.int/web/gaia/dpac/consortium}). Funding for the DPAC has been provided by national institutions, in particular the institutions participating in the {\it Gaia} Multilateral Agreement.

The authors wish to recognize the significant contributions of David Doss to McDonald Observatory on the occasion of his retirement; his tireless efforts throughout his career enabled this research as well as that of multiple generations of white dwarf astronomers. 

The authors thank D.~Winget and M.~H.~Montgomery for helpful discussions after the initial discovery of this object's variability.  The authors also thank J.\ Farihi, P.\ Dufour, and M.\ Kilic for insightful discussions during the preparation of this manuscript.  We are also grateful for the time and efforts of J.~Kuhne, B.~Roman, and the rest of the staff at McDonald Observatory for enabling remote observing with the Otto Struve Telescope during the COVID-19 pandemic.

This material is based upon work supported by the National Science Foundation under Grant No. AST-1910551.

\end{acknowledgements}

\bibliography{EGGR_156}{}
\bibliographystyle{aasjournal}

%

\end{document}